\documentclass{article}
\usepackage{frascatiphys}
\usepackage{graphicx}

\usepackage{amsmath}
\usepackage{amsfonts}

\newcommand{\beq}{\begin{equation}}
\newcommand{\eeq}{\end{equation}}
\newcommand{\beqa}{\begin{eqnarray}}
\newcommand{\eeqa}{\end{eqnarray}}
\def\lsim{\raise0.3ex\hbox{$<$\kern-0.75em\raise-1.1ex\hbox{$\sim$}}}
\def\gsim{\raise0.3ex\hbox{$>$\kern-0.75em\raise-1.1ex\hbox{$\sim$}}}
\def\kk{{\kappa}}
\def\pp{{\hat{p}}}

\begin{document}
\title{ 
HEAVY FLAVOUR IN HIGH-ENERGY NUCLEAR COLLISIONS:\\ THEORY OVERVIEW OF TRANSPORT CALCULATIONS}
\author{
Andrea Beraudo        \\
{\em INFN - Sezione di Torino}
}
\maketitle
\baselineskip=10pt
\begin{abstract}
Transport calculations are the tool to study medium modifications of heavy-flavour particle distributions in high-energy nuclear collisions. We give a brief overview on their state-of-the art, on the information one can extract, on the questions remaining open and on further analysis to carry out in the near future.
\end{abstract}
\baselineskip=14pt

\section{Introduction}
The description of heavy-flavour (HF) observables in relativistic heavy-ion collisions requires to develop an involved multi-step setup. One has to simulate the initial $Q\overline{Q}$ production in a hard process and for this automated QCD event generators -- to validate against proton-proton data -- are available, possibly supplemented with initial-state effects, like nuclear Parton Distribution Functions. One has then to rely on a description of the background medium, provided by hydrodynamic calculations tuned to reproduce soft-hadron data. One needs then to model the interaction of the heavy quarks with the medium, summarized in a few transport coefficients in principle derived from QCD, but for which we are still far from a definite answer for the experimentally relevant conditions. One can then describe the heavy-quark dynamics in the medium: under well-defined kinematic conditions this can be done rigorously through transport equations, which however require the above transport coefficients as an input. The heavy quarks, once they reach a fluid-cell below the deconfinement temperature, undergo hadronization; one can expect that, at variance with the case of elementary collisions, some sort of recombination with the abundant nearby partons is at work. This represents an item of interest in itself, but at the same time the related theoretical uncertainties affect the predictions for the final hadronic observables, preventing one from getting an unambiguous information on the partonic stage. Finally, $D$ and $B$ mesons can still rescatter with the surrounding hadrons before reaching kinetic freeze-out: also this possible effect deserves some study.
In the following we will discuss the above items in a more quantitative way, eventually considering how the above setup can be applied to perform more refined analysis of the final HF particle distributions and extended to the study of HF production in small systems, like high-multiplicity proton-proton/nucleus collisions.

\section{Transport calculations: theoretical setup}
\begin{figure}[!ht]
\begin{center}
\includegraphics[clip,height=5cm]{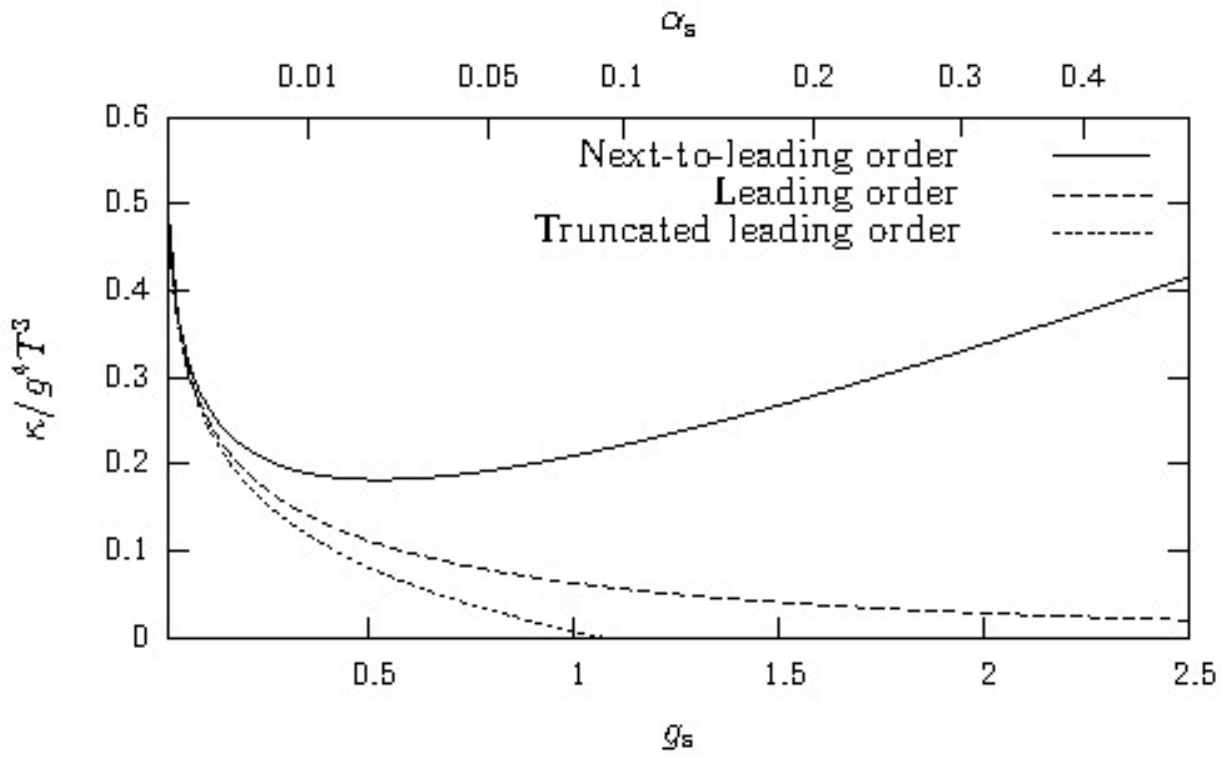}
\includegraphics[clip,height=5cm]{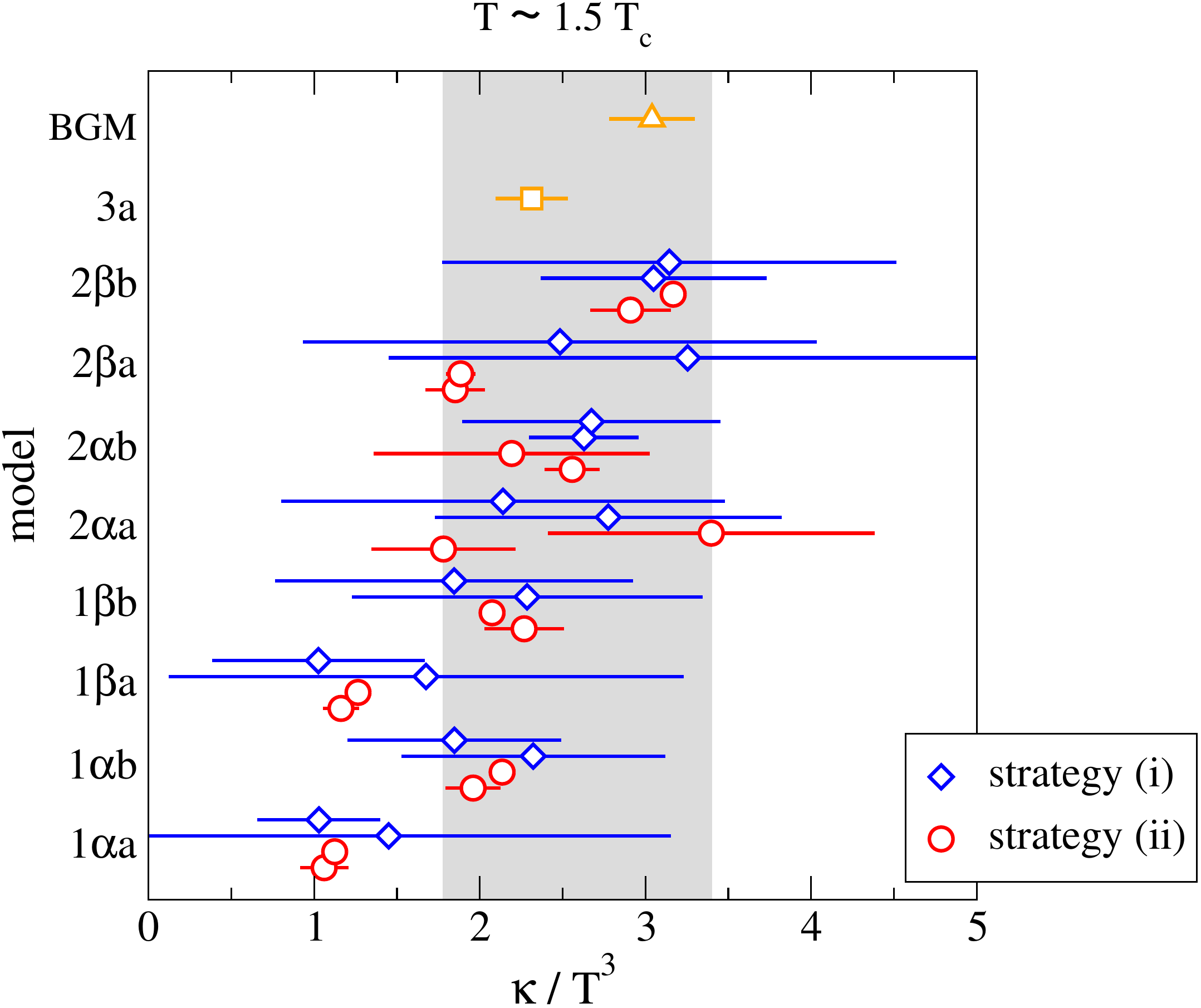}
\caption{Momentum-diffusion coefficient $k$. Left panel: result of a NLO weak-coupling calculation~\cite{CaronHuot:2008uh}. Right panel: continuum-extrapolated lattice-QCD results for a gluon plasma~\cite{Francis:2015daa}.}\label{fig:transp-coeff} 
\end{center}
\end{figure}
Transport calculations for $c$ and $b$ quarks in heavy-ion collisions are usually formulated in terms of the relativistic Langevin equation~\cite{Moore:2004tg,Alberico:2011zy,He:2013zua}
\beq
{\Delta \vec{p}}/{\Delta t}=-{\eta_D(p)\vec{p}}+{\vec\xi(t)}.\label{eq:Langevin}
\eeq
In the time-step $\Delta t$ the heavy-quark momentum changes due to a deterministic friction force (quantified by the coefficient $\eta_D$) and a random noise term $\vec\xi$ fixed by its temporal correlator
\beq
\langle\xi^i(\vec p_t)\xi^j(\vec p_{t'})\rangle\!=\!{b^{ij}(\vec p_t)}{\delta_{tt'}}/{\Delta t},\quad{\rm with}\quad{b^{ij}(\vec p)}\!\equiv\!{\kk_\|(p)}\pp^i\pp^j+{\kk_\perp(p)}(\delta^{ij}\!-\!\pp^i\pp^j).
\eeq
In the above the transport coefficients $\kk_\|(p)$ and $\kk_\perp(p)$ describe the longitudinal and transverse momentum broadening acquired by the heavy quark while propagating through the medium. In the non-relativistic limit one can ignore the dependence on the heavy-quark momentum and simply set $\kk_\|(p)\!=\!\kk_\perp(p)\!\equiv\!\kappa$. In such a limit, from the large-time behaviour of the average squared displacement, one can identify the spatial diffusion coefficient $D_s$:
\beq
\langle\vec x^2(t)\rangle\underset{t\to\infty}{\sim}6{D_s}t\quad{\rm with}\quad {D_S=\frac{2T^2}{\kappa}}.\label{eq:Ds}
\eeq
The latter is often used to quantify the strength of the coupling with the medium. 

First principle theoretical results are available for $\kappa$ in the static $M\!\to\!\infty$ limit, arising both from analytic weak-coupling calculations~\cite{CaronHuot:2008uh} and from lattice-QCD simulations~\cite{Francis:2015daa}. In the left panel of Fig.~\ref{fig:transp-coeff} one can see how, for realistic values of $\alpha_s$, the weak-coupling calculation for $\kappa$ receives large NLO corrections, arising mainly from overlapping scattering processes with the light partons from the medium. In the case of lattice-QCD calculations, on the other hand, the right panel of Fig.~\ref{fig:transp-coeff} shows that the final result is affected by large systematic theoretical uncertainties arising from the extraction of real-time information from simulations performed in an Euclidean spacetime. Taking into account the large systematic uncertainties, the two calculations provide results in rough agreement.

\begin{figure}[!ht]
\begin{center}
\includegraphics[clip,height=5cm]{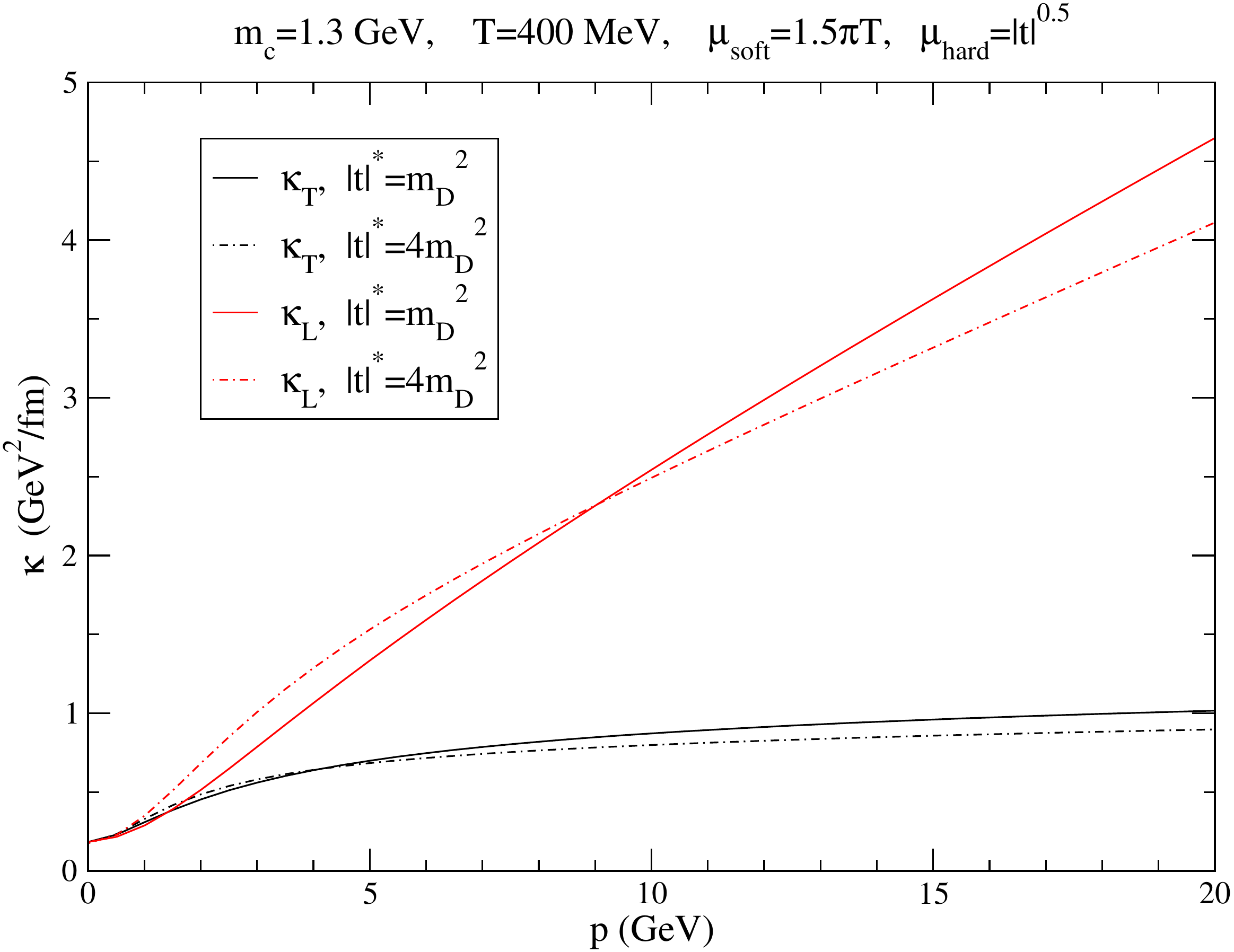}
\includegraphics[clip,height=4.6cm]{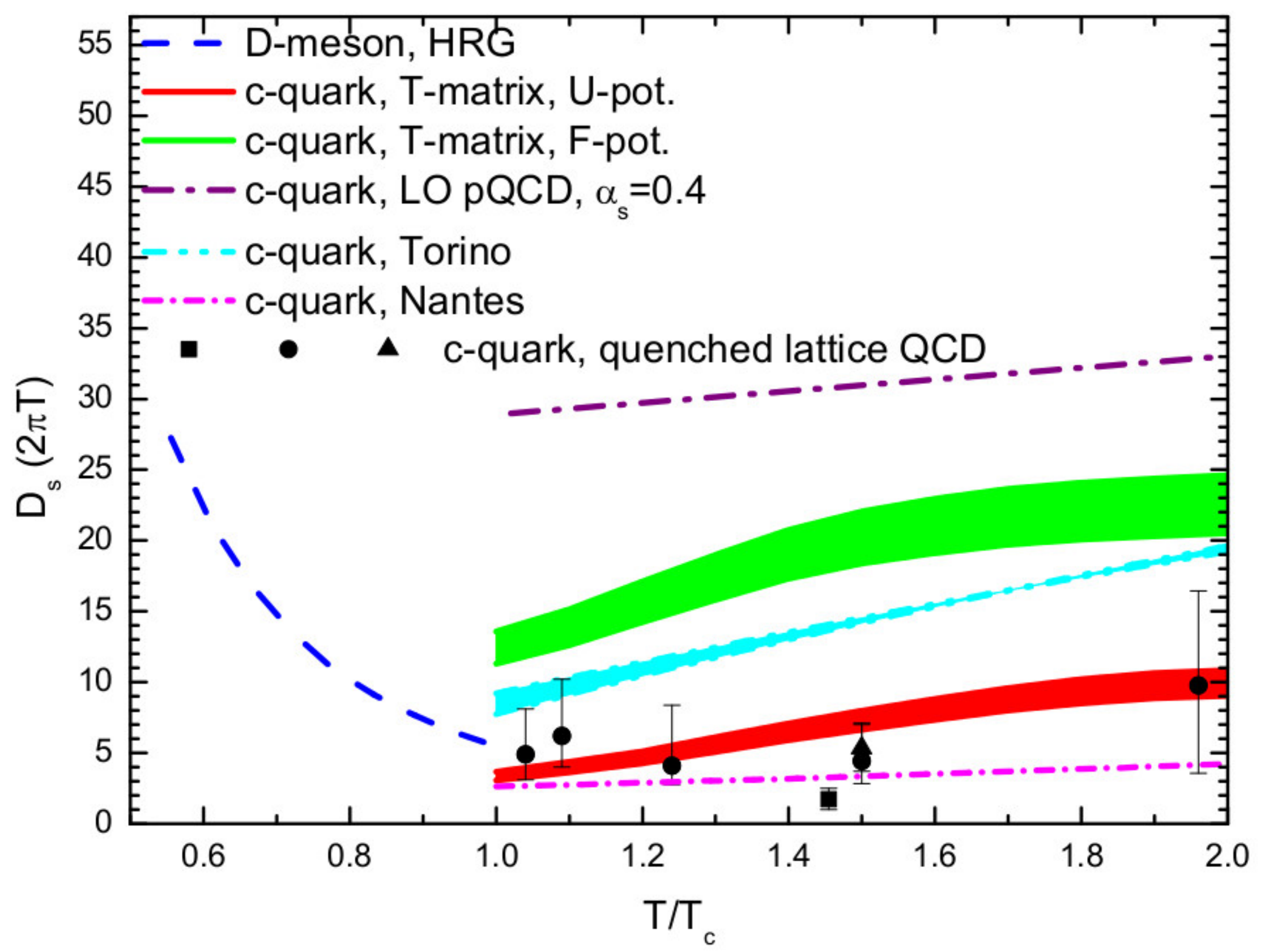}
\caption{Left panel: charm momentum-broadening coefficients from the weak-coupling calculation in~\cite{Alberico:2013bza}. Right panel: charm spatial diffusion coefficient provided by different theoretical models~\cite{Prino:2016cni}.}\label{fig:transp-coeff2} 
\end{center}
\end{figure}
Unfortunately, experimentally, the accessible kinematic range for heavy-flavour hadron detection covers mainly relativistic momenta. Hence, transport simulations require a theoretical input going beyond the result of the above static calculations and one needs to consider the full momentum dependence of the coefficients. This was done for instance in~\cite{Alberico:2013bza} with a weak-coupling calculation with Hard-Thermal-Loop (HTL) resummation of medium effects: results for $\kk_\|(p)$ and $\kk_\perp(p)$ are displayed in the left panel of Fig.~\ref{fig:transp-coeff2}. Although the figure clearly shows that the dependence on the particle momentum is relevant and very different for the transverse and the longitudinal coefficients, in comparing the results of different transport calculations one very often simply employs the spatial diffusion coefficient $D_s$ defined in Eq.~(\ref{eq:Ds}) to enlighten the differences of the various models; a collection of results for $D_s$ obtained under different theoretical frameworks is shown in the right panel of Fig.~\ref{fig:transp-coeff2}~\cite{Prino:2016cni}.

\section{Heavy-flavour production in nucleus-nucleus collisions}
\begin{figure}[!ht]
\begin{center}
\includegraphics[clip,width=0.48\textwidth]{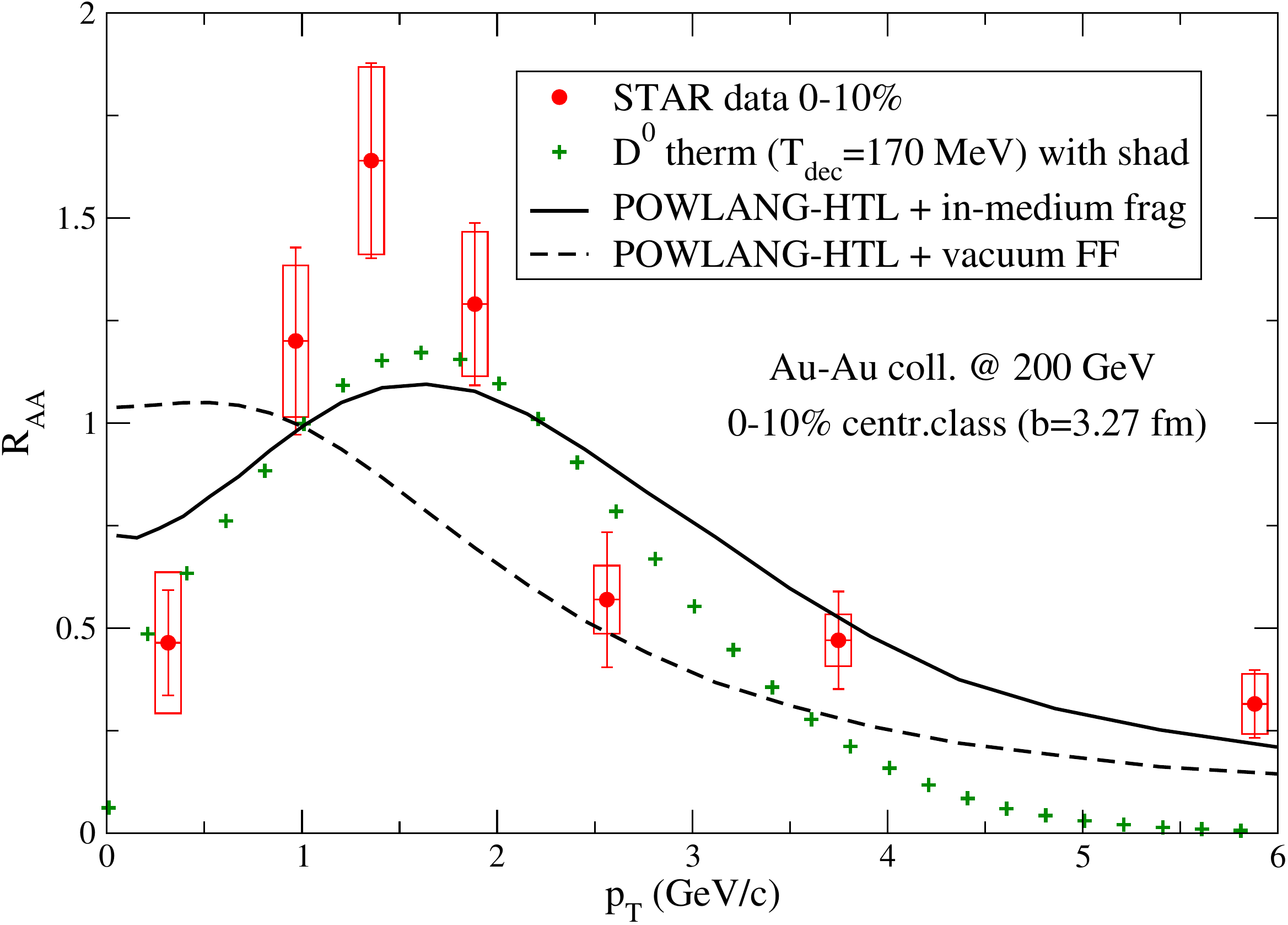}
\includegraphics[clip,width=0.48\textwidth]{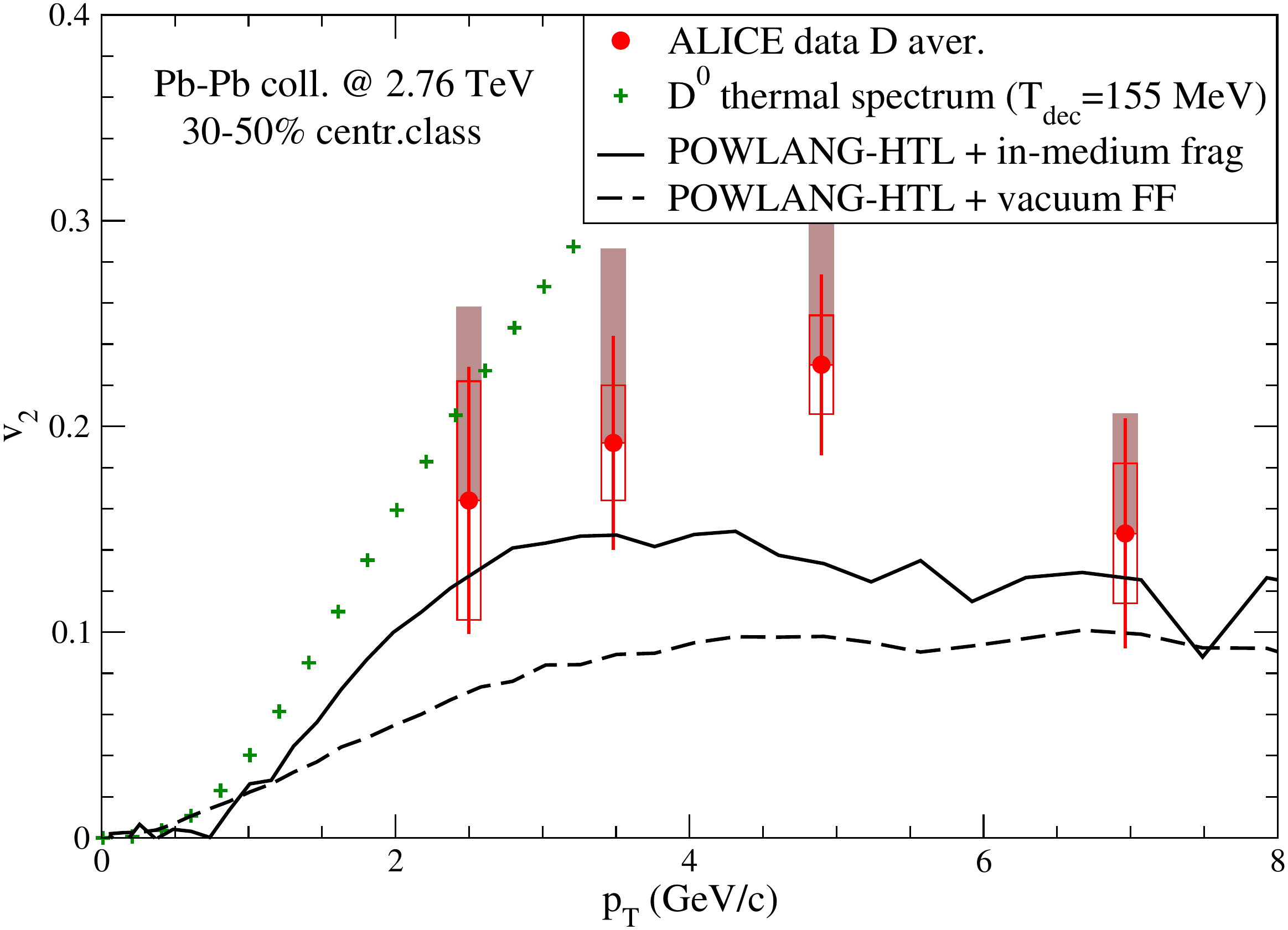}
\caption{Results for the $D$-meson $R_{\rm AA}$ and $v_2$ in nucleus-nucleus collisions~\cite{Beraudo:2014boa} arising from the heavy quark transport (dashed curves) and transport + in-medium hadronization (continuous curves) compared to experimental data~\cite{Adamczyk:2014uip,Abelev:2013lca}. Green crosses refer to the kinetic-equilibrium limit.}\label{fig:POWLANG} 
\end{center}
\end{figure}
Eq.~(\ref{eq:Langevin}), once interfaced with a realistic model for the evolution of the background medium, allows one to study the propagation of the heavy quarks throughout the fireball produced in high-energy nuclear collisions until they reach a fluid cell below the critical deconfinement temperature, where they are forced to hadronize. One can then study the modification of the HF hadron distributions introduced by the medium. This is usually done via the nuclear modification factor $R_{\rm AA}$, i.e. the ratio of the spectra in A+A and p+p collisions (rescaled by the number of binary nucleon-nucleon collisions), and the $v_n\equiv\langle\cos[n(\phi-\psi_{n})]\rangle$ coefficients, quantifying the asymmetry of the azimuthal particle distributions. One usually focus on the elliptic-flow coefficient $v_2$, arising mainly from the finite impact parameter of the A+A collision: the initial geometric deformation is converted by the pressure gradients into an anisotropic flow of the matter. The interaction of the heavy quarks with the medium, depending on its strength, leads to a quenching of the HF hadron spectra at high $p_T$ -- due to parton energy-loss -- to a possible enhancement in the intermediate $p_T$ region -- partly due to the radial flow inherited from the fireball and partly due to the conservation of the number of charm/beauty quarks -- and to an asymmetry in the azimuthal distribution of HF particles, reflecting the anisotropic geometry and flow of the background medium.  
Concerning the effect of hadronization, while the latter in the vacuum is usually described in terms of fragmentation functions and leads to an energy degradation of the parent parton, there is evidence that in nuclear collisions recombination with the nearby partons from the medium plays a major role. The process was modeled in several ways in the literature~\cite{Greco:2003vf,Gossiaux:2009mk,Beraudo:2014boa}, leading however to similar effects in the particle distributions. In Fig.~\ref{fig:POWLANG} we display some results of the POWLANG model~\cite{Beraudo:2014boa}, in which hadronization is modeled via fragmentation of color-singlet string/clusters formed through the recombination of a heavy quark with a light thermal anti-quark from the same fluid-cell. It turns out that the light parton from the medium transfers part of its collective (radial and elliptic) flow to the final $D$-meson, leading to a bump at moderate $p_T$ in the $R_{\rm AA}$ and to an enhancement of the $v_2$ and moving theory predictions closer to the experimental data~\cite{Adamczyk:2014uip,Abelev:2013lca}.

\begin{figure}[!ht]
\begin{center}
\includegraphics[clip,width=0.3\textwidth]{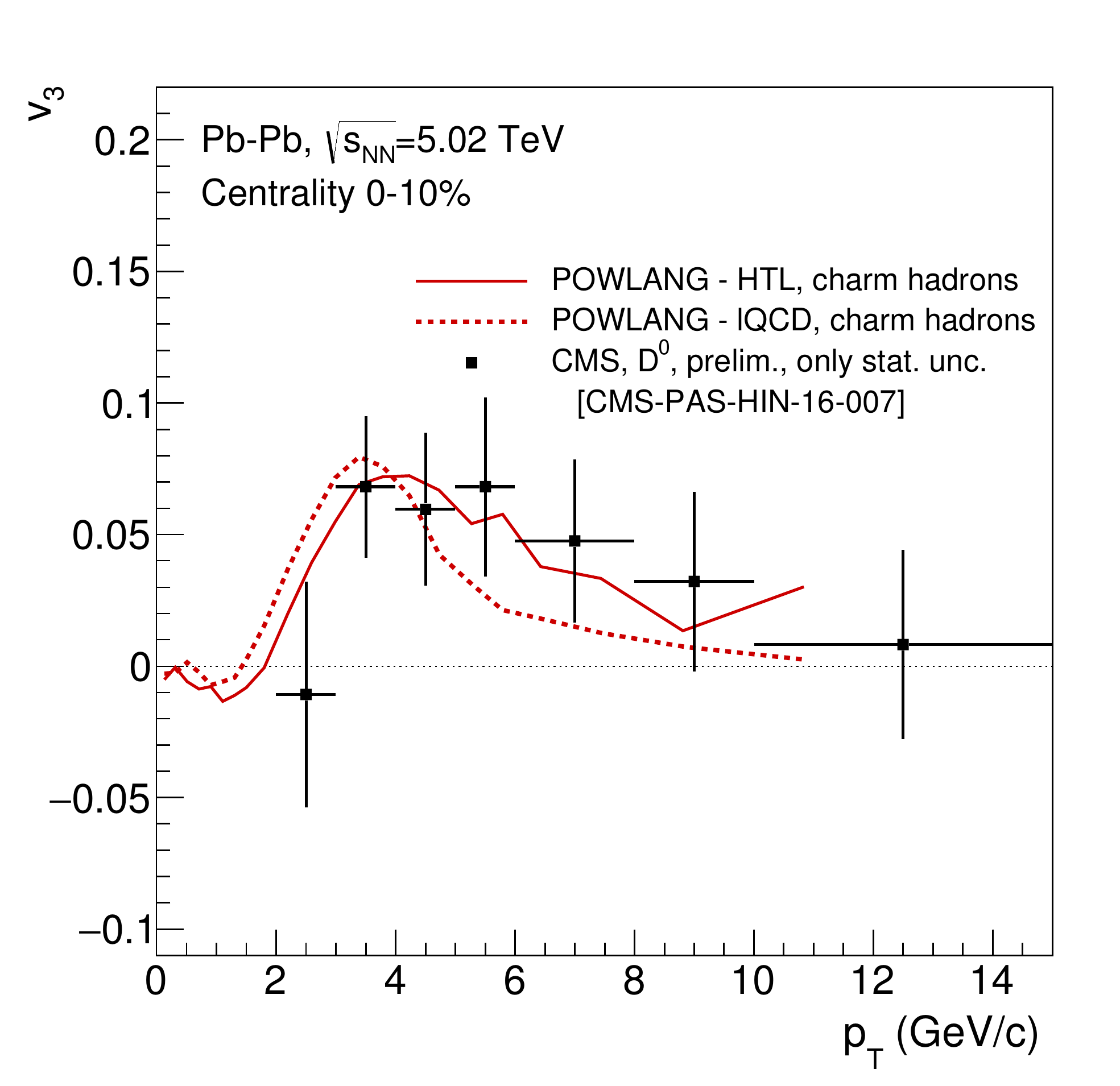}
\includegraphics[clip,width=0.3\textwidth]{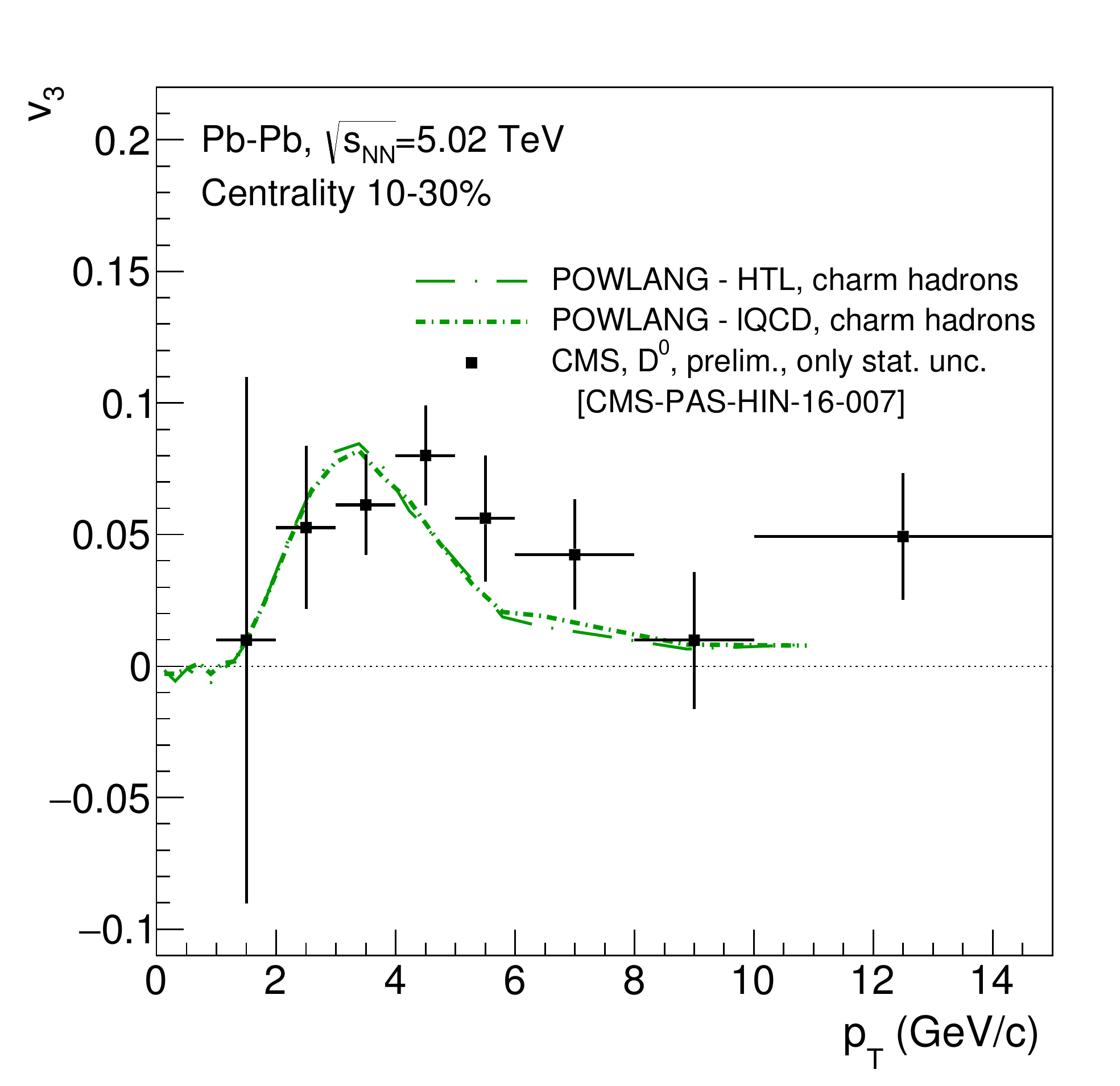}
\includegraphics[clip,width=0.3\textwidth]{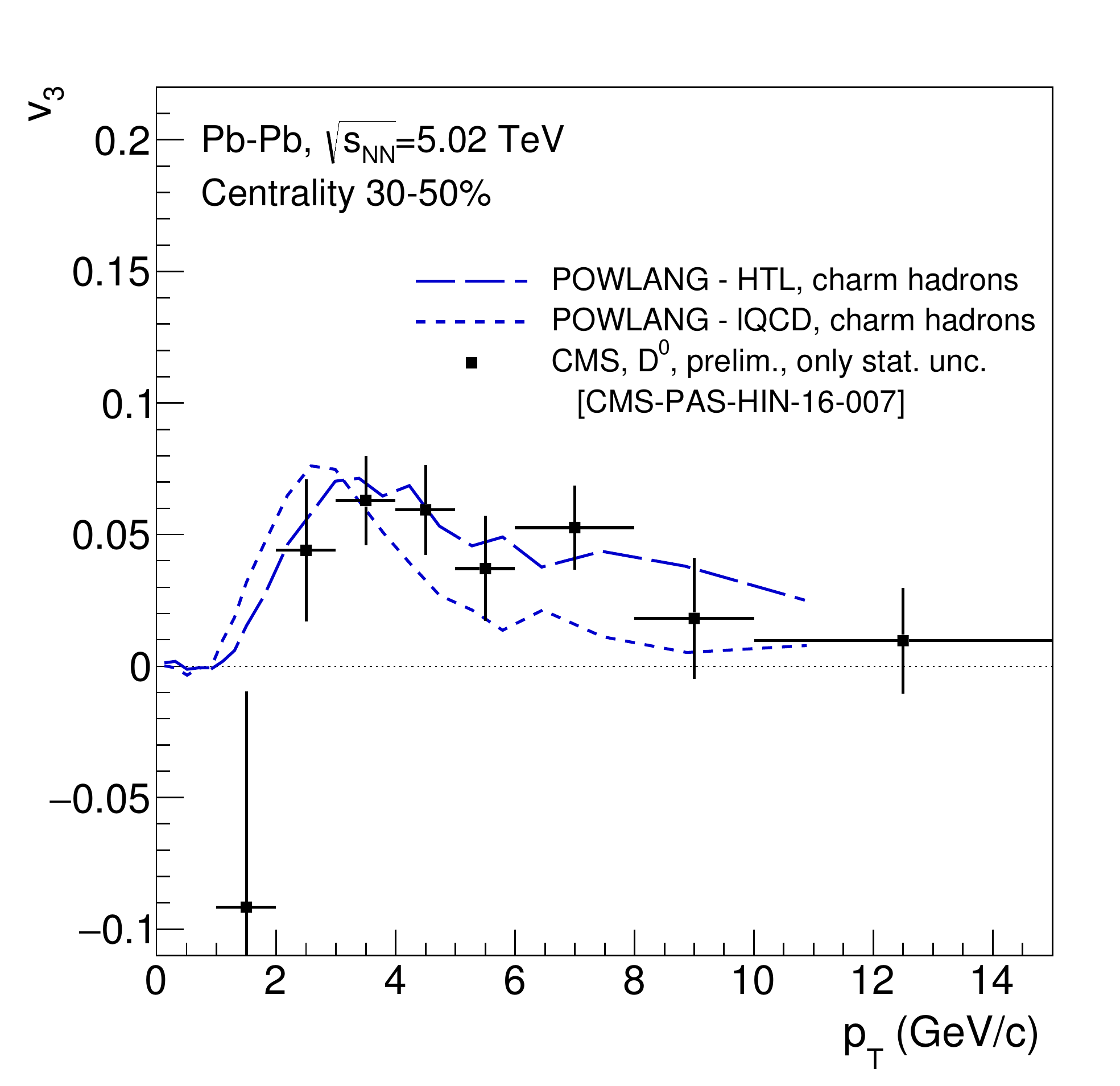}
\caption{POWLANG predictions for the triangular flow of D mesons in Pb-Pb collisions at the LHC~\cite{Beraudo:2017gxw} compared to preliminary CMS data~\cite{Sirunyan:2017plt}.}\label{fig:v3} 
\end{center}
\end{figure}
Besides the initial elliptic deformation of the fireball -- arising mainly from the finite impact parameter of the collision of the two nuclei -- event-by-event fluctuations in the nucleon positions and, possibly, of sub-nucleonic degrees of freedom represent a further source of azimuthal asymmetry, giving rises to higher flow harmonics, absent in the case of smooth initial conditions. At this regard, in Fig.~\ref{fig:v3} we display the predictions of the POWLANG model for the triangular flow $v_3$ of D mesons in Pb-Pb collisions at $\sqrt{s_{NN}}\!=\!5.02$ TeV~\cite{Beraudo:2017gxw} compared to preliminary CMS data~\cite{Sirunyan:2017plt}.

\begin{figure}[!ht]
\begin{center}
\includegraphics[clip,height=5.5cm]{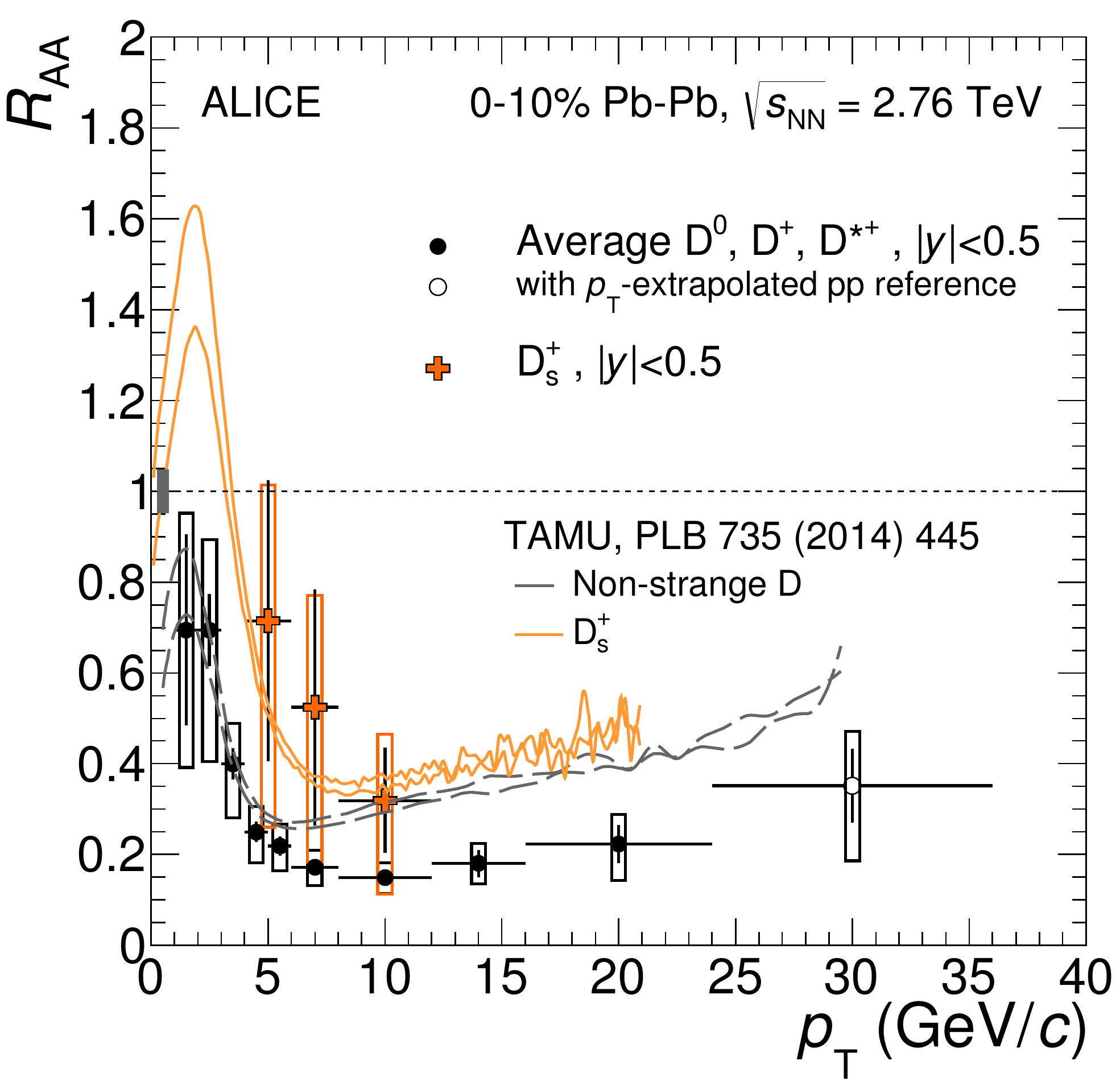}
\includegraphics[clip,height=5.5cm]{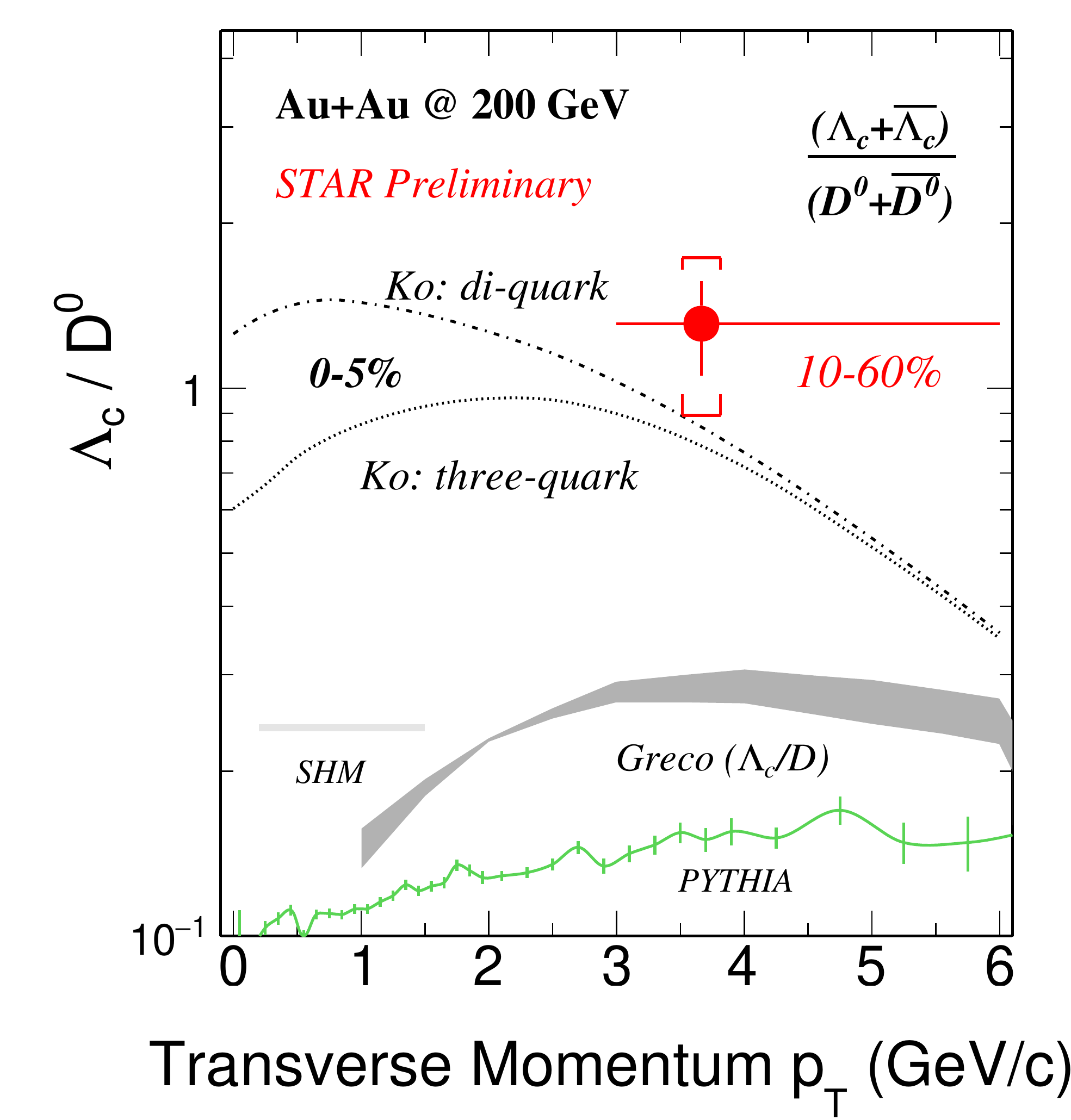}
\caption{First results for $D_s$~\cite{Adam:2015jda} and $\Lambda_c$~\cite{Zhou:2017ikn} production in nucleus-nucleus collisions compared to transport calculations including in-medium hadronization~\cite{He:2014cla,Oh:2009zj}.}\label{fig:DsLambdac} 
\end{center}
\end{figure}
Besides modifying the momentum distribution of the final hadrons, recombination can also change the heavy-flavour hadrochemistry, leading for instance to an enhanced production of $D_s$ mesons and $\Lambda_c$ baryons. Predictions obtained with models based on the formation of resonant states around the phase transition~\cite{He:2014cla} and on the coalescence of quarks and di-quarks~\cite{Oh:2009zj} are shown in Fig.~\ref{fig:DsLambdac} and compared to ALICE~\cite{Adam:2015jda} and preliminary STAR data~\cite{Zhou:2017ikn}.

\begin{figure}[!ht]
\begin{center}
\includegraphics[clip,height=5.5cm]{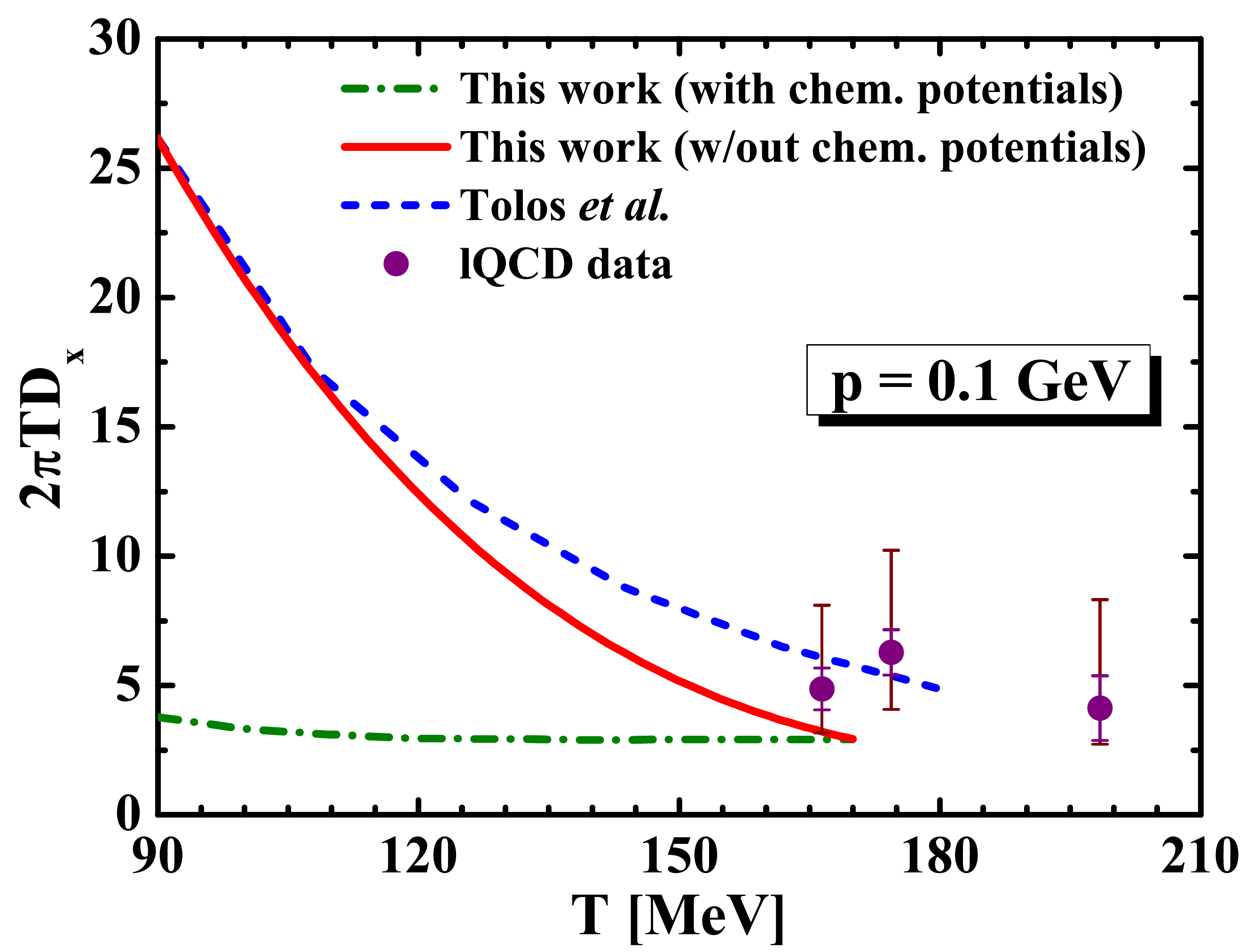}
\includegraphics[clip,height=5.5cm]{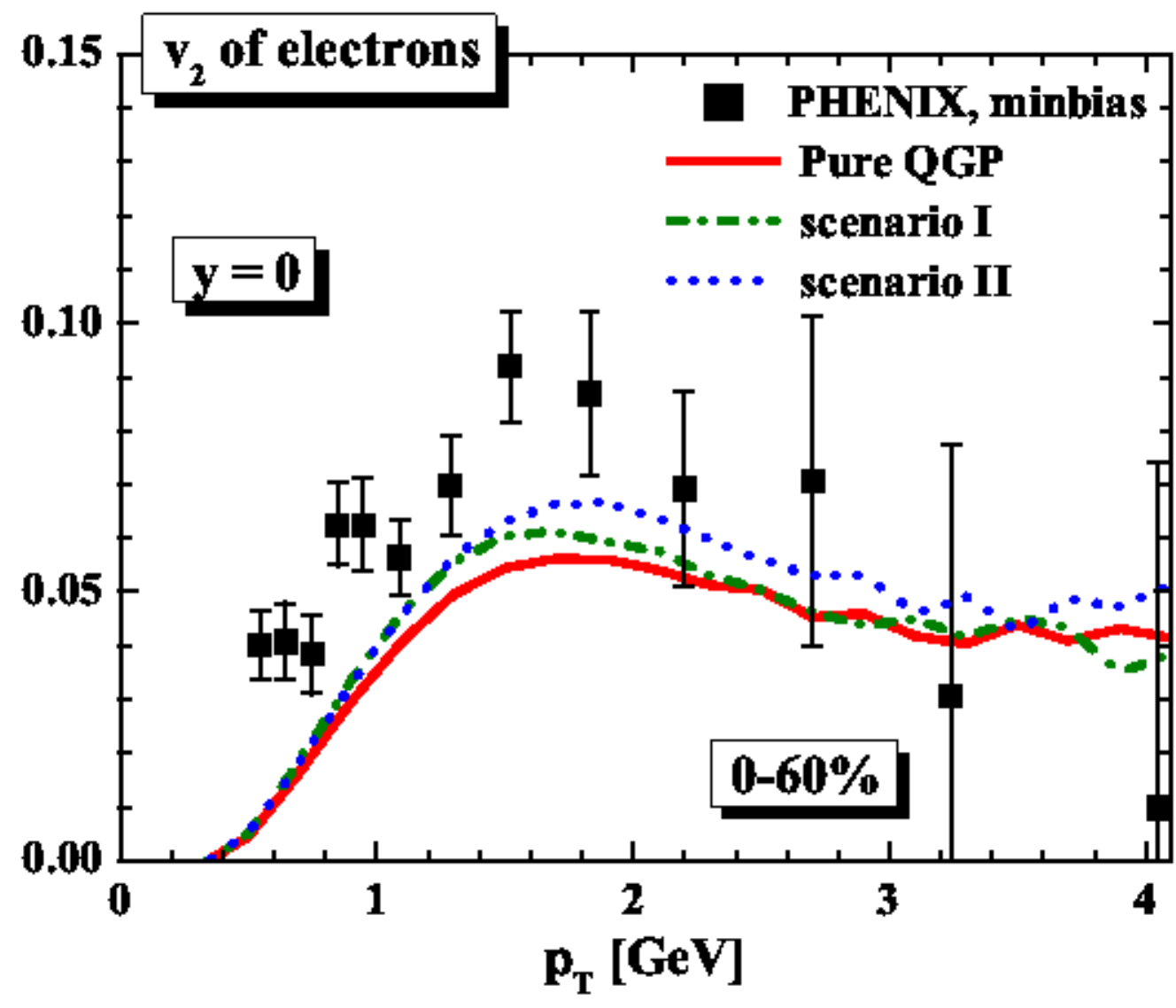}
\caption{The heavy-quark spatial diffusion coefficient in the hadronic phase (left panel) and its effect on a transport calculation~\cite{Ozvenchuk:2014rpa} (right panel). Data refer to the heavy-flavour decay electron $v_2$ measured in~\cite{Adare:2006nq}.}\label{fig:hadronic} 
\end{center}
\end{figure}
Finally, although crossing a medium with a lower temperature and hence milder values of the transport coefficients, heavy-flavour particles can suffer rescattering also in the hadronic phase, where the radial and elliptic flow of the fireball is the largest. It is then of interest to evaluate within some effective chiral Lagrangian the transport coefficients of $D/B$-mesons in a gas of light hadrons~\cite{Ozvenchuk:2014rpa}, whose values -- as a function of the temperature -- turn out to join quite smoothly the results in the partonic phase (left panel of Fig.~\ref{fig:hadronic}). One can then include also the possibility of rescattering in the hadronic phase in the transport simulations, however the effect on the final observables is found to be quite small, as shown in the right panel of Fig.~\ref{fig:hadronic} which refers to the elliptic flow of electrons from semi-leptonic decays of charm and beauty hadrons~\cite{Adare:2006nq}.

\section{Heavy-flavour production in small systems}
\begin{figure}[!ht]
\begin{center}
\includegraphics[clip,width=0.48\textwidth]{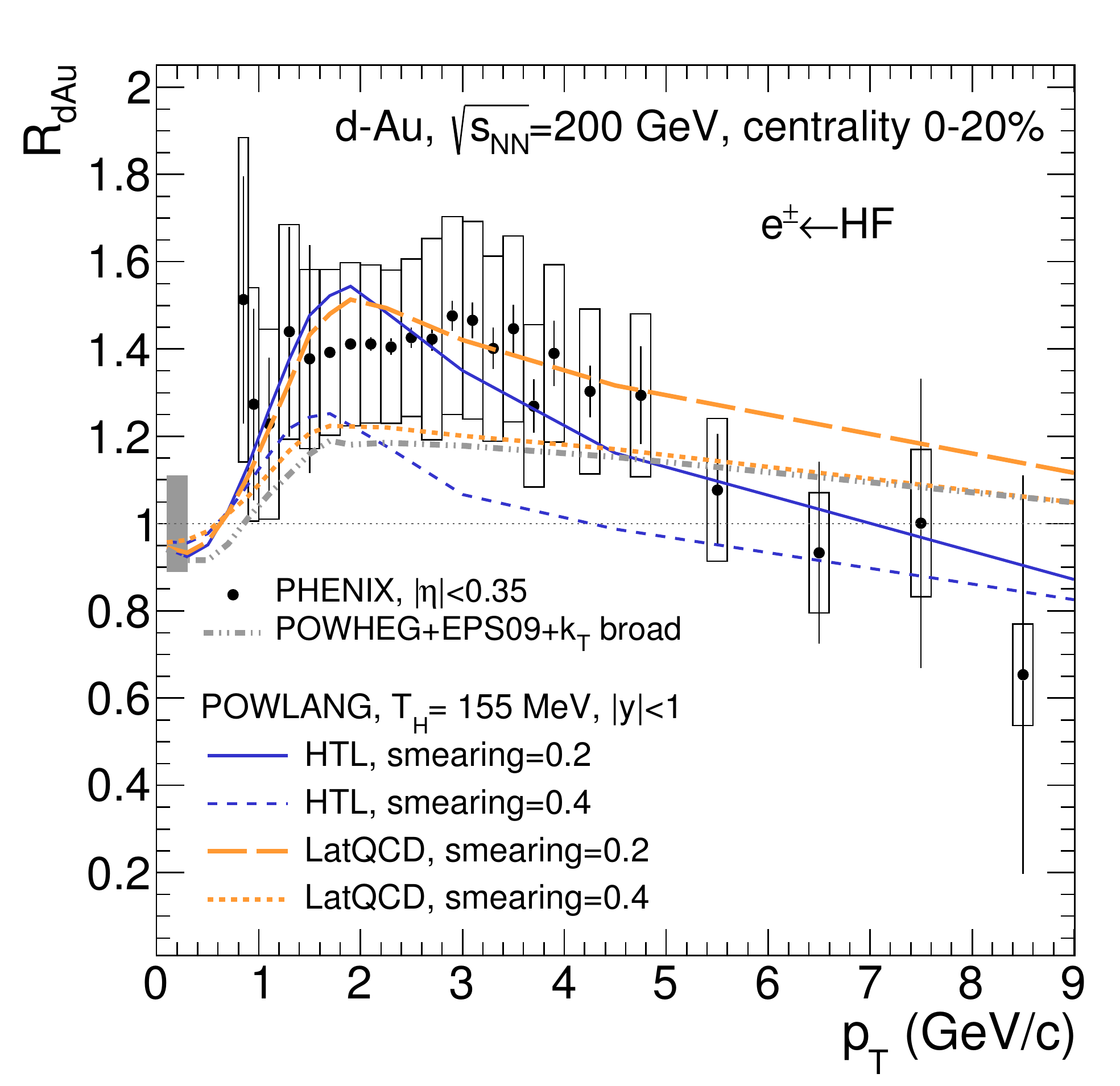}
\includegraphics[clip,width=0.48\textwidth]{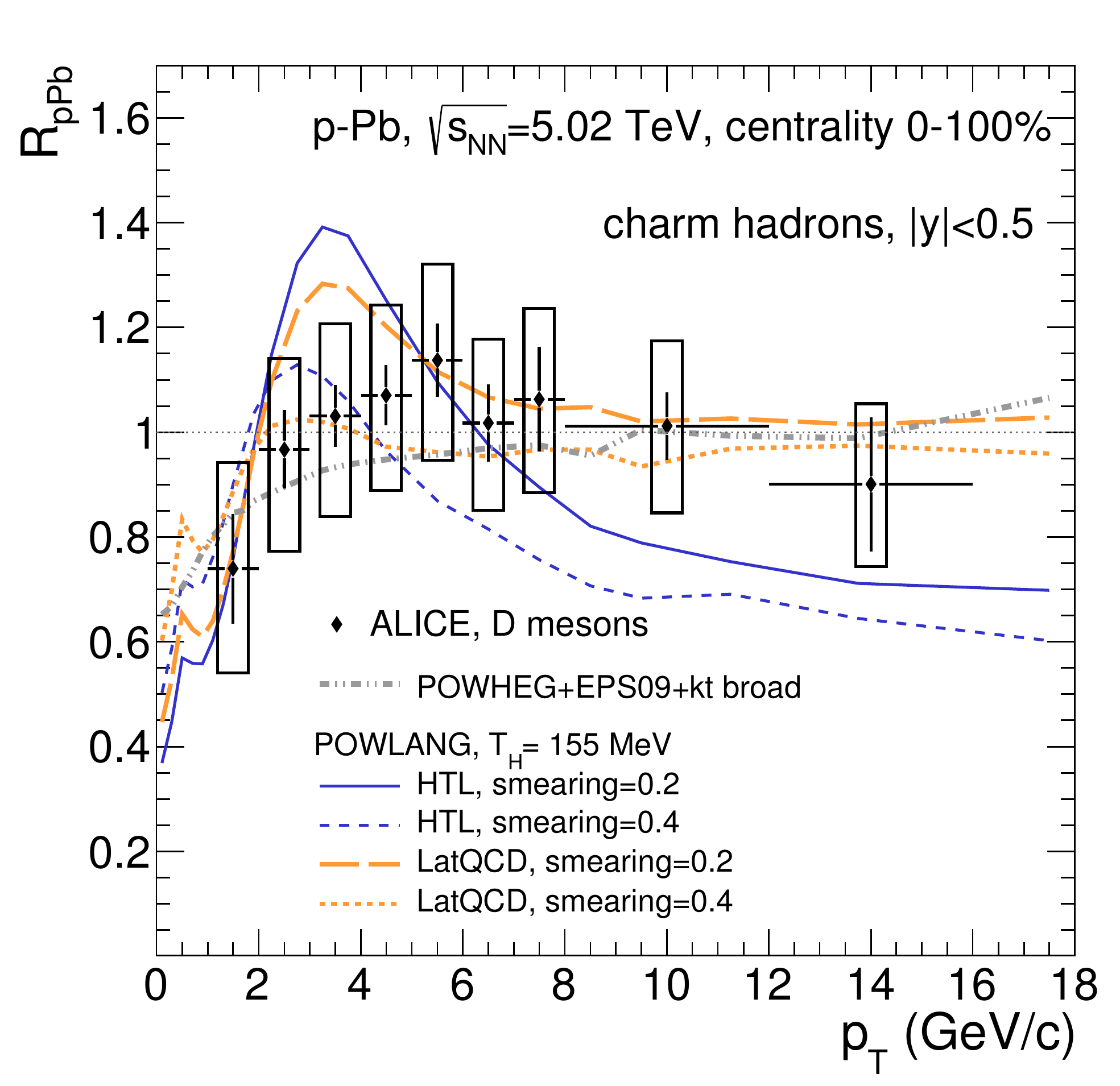}
\caption{The nuclear modification factor of heavy-flavour decay electrons and $D$-mesons in d-Au and p-Pb collisions at RHIC and LHC. Transport-model results with different transport coefficients~\cite{Beraudo:2015wsd} are compared to experimental data~\cite{Adare:2012yxa,Abelev:2014hha}.}\label{fig:small} 
\end{center}
\end{figure}
The observation of signatures of collective effects in measurements of particle correlations performed in high-multiplicity proton-nucleus and proton-proton collisions as well as the universal trend of the integrated particle yields as a function of the final hadron multiplicity led people to wonder about the nature of the possible medium produced in such events. In parallel, no strong evidence of jet-quenching was found in p-p and p-A data: the medium seems to be strongly interacting but not very opaque to highly energetic probes. Hence the interest in studying heavy-flavour production within a transport setup also in the case of small systems. This was done for instance by the POWLANG authors in~\cite{Beraudo:2015wsd}. Results are shown in Fig.~\ref{fig:small}, referring to electrons from charm and beauty decays in central d-Au collisions and to charmed hadrons in minimum bias Pb-Pb events. Deviations from unity arise from the interplay of several effects: nuclear PDF's, $k_T$-broadening in cold nuclear matter, transport in the partonic phase and in-medium hadronization. The large systematic uncertainties of the experimental data~\cite{Adare:2012yxa,Abelev:2014hha} do not allow one to draw firm conclusions, but leave room for medium effects. 

\section{Conclusions and perspectives}
The comparison of current transport calculations with experimental data provides strong evidence that charm quarks interact significantly with the medium formed in heavy-ion collision, which affects both their propagation in the plasma and
their hadronization. A number of experimental challenges or theoretical questions remain to be answered: charm measurements down to $p_T\to 0$ will provide more solid information on its possible thermalization and production cross-section (of relevance to quantify charmonium suppression!); $D_s$ and $\Lambda_c$ measurements will shed light on the issue of possible changes in the HF hadrochemistry and, again, on the total charm cross-section; finally, beauty measurements -- due to the large heavy-quark mass -- will be the golden channel to extract information on the HF transport coefficient from the data.

\end{document}